\documentclass[11pt]{article}

\usepackage{arxiv}
\usepackage{amsmath}
\usepackage{amssymb}
\usepackage{booktabs}
\usepackage{graphicx}
\usepackage{hyperref}
\usepackage[numbers,sort&compress]{natbib}
\usepackage{placeins}

\title{Wavelet as Tokenizer: Preliminary Results on a Shared Wavelet Token Schema for Natural Signals}

\author{Shenghao Ding\\
Yet Another AI\\
\texttt{shenghao.ding@yetanother.ai}}

\renewcommand{\headeright}{}
\renewcommand{\undertitle}{}
\renewcommand{\shorttitle}{Wavelet as Tokenizer}

\date{}

\begin{document}

\maketitle

\begin{abstract}
This paper studies whether audio, images, and video can share a common wavelet
token schema rather than relying on separate modality-specific latent grids. It
introduces a continuous-token model built around a one-level Haar
DWT/IDWT frontend, a shared coefficient-token layout, optional structural
metadata, lightweight modality value adapters, and a shared token-wise
encoder-decoder trunk. On Speech Commands, EuroSAT RGB, and DAVIS 2017 data, a
dense shared model reaches 39.92 dB audio, 29.37 dB image, and 23.93 dB video
PSNR. A matched-rate sweep under continuous latent scalar budgets indicates that
the visual gains are not explained solely by latent capacity, while also showing
that additive metadata embeddings are not a universal source of improvement.
Finally, fixed-rate energy selection provides an effective non-parametric baseline:
\texttt{energy\_global} improves average PSNR over uniform selection by 16.73 dB
for audio, 16.90 dB for images, and 15.86 dB for video under compressed keep
ratios. Masked sparse training reaches 34.45 dB video PSNR with 50\% of dense
tokens. The results support a unified wavelet token schema and sparse token
interface, while stopping short of establishing a universal discrete vocabulary.
\end{abstract}

\section{Introduction}

Tokenizers have become the interface between high-dimensional data and modern
generative models. For text, tokenization is largely symbolic: a sequence of
characters or words is mapped to a sequence of discrete ids. For natural signals,
however, tokenization is closer to lossy compression. An audio codec, an image
autoencoder, and a video tokenizer must all decide which parts of a dense signal
deserve representation under a limited token budget, while preserving the
perceptual and semantic information needed by downstream models.

Most existing tokenizers are designed around the geometry of a single modality.
Images are often divided into patches or encoded into dense latent grids; audio
tokenizers compress one-dimensional waveforms or time-frequency features; video
tokenizers extend visual latents over time. These designs have produced strong
modality-specific systems, but they leave open whether audio, images, and video
can share a common token language rather than merely sharing a neural backbone.

This paper studies \emph{Wavelet as Tokenizer} (WAT), a multi-scale
tokenization framework for natural signals. Each modality is viewed as a
sampled or parameterized field $x: \Omega \rightarrow \mathbb{R}^C$, where $\Omega$ may be
time for audio, a two-dimensional spatial domain for images, or spacetime for
video. A wavelet or learned lifting transform maps the field into localized
multi-scale coefficients. Tokens are then formed from coefficient blocks together
with their structural metadata:
\[
(\text{value}, \text{modality}, \text{rank}, \text{scale}, \text{location},
\text{subband}).
\]
In this view, a token is not only a latent vector. It is a compact statement
about what kind of signal variation occurred, at which scale, in which subband,
and where in the underlying domain. Discrete code ids can later be added on top
of this schema, but this paper deliberately studies the continuous case first.

The motivation is not simply to use wavelets as another downsampling layer.
Recent visual tokenizers, including joint image-video systems and wavelet-space
tokenizers such as Cosmos, show that Haar wavelet front-ends and neural
autoencoders can be highly effective for images and videos
\citep{wang2024omnitokenizer,nvidia2025cosmos}. The central question is whether
wavelet coefficients can serve as a modality-agnostic token substrate for
one-dimensional audio, two-dimensional images, and three-dimensional spacetime
video, with a future path toward continuous 3D fields.

This framing leads to two design commitments. First, audio is treated as a
first-class modality rather than an afterthought to a visual tokenizer. Second,
the scale, subband, and location of each coefficient are exposed to the token
model instead of being hidden inside a dense latent grid. These commitments make
it possible to compare fixed-rate dense tokenization with adaptive sparse token
allocation, where smooth regions can be represented by fewer coarse tokens while
edges, transients, textures, and motion receive more fine-scale tokens.

The present study reports a dense, one-level Haar formulation with continuous
latent tokens and no codebook. The experiments focus on a narrower question than
the full long-term agenda: whether a single shared token trunk can reconstruct
real 1D audio, 2D images, and 3D video from the same wavelet token grammar. The
results answer this question positively for small-scale autoencoding, identify
audio coefficient scaling as an important normalization issue, and leave
discrete quantization, learned sparse allocation, and downstream generative
modeling to future work.

\subsection{Contributions}

This paper makes four contributions. First, it defines a shared wavelet
coefficient token schema for audio, images, and video, with explicit modality,
rank, scale, subband, and position fields. Second, it evaluates a compact
continuous autoencoder that processes the three modalities through a shared
token trunk while retaining only lightweight modality-specific value adapters and
inverse-wavelet reconstruction paths. Third, it compares shared and separate
models under matched continuous latent scalar budgets, showing that the visual
advantage of the shared schema is not explained solely by bottleneck size in
these controlled experiments. Fourth, it evaluates fixed-rate energy token
selection and masked sparse training, providing evidence that wavelet
coefficient energy is a useful cross-modal allocation signal.

\section{Background and Related Work}

\paragraph{Tokenization as signal compression.}
Tokenizers for natural signals sit between classical compression and learned
representation learning. A tokenizer must reduce a high-rate signal to a compact
sequence while preserving the information that matters for reconstruction,
generation, or downstream prediction. This differs from text tokenization, where
the input is already symbolic. For audio, images, video, and 3D scenes, the
tokenizer must learn or choose a sampling basis, allocate bits or tokens across
the signal, and expose a representation that can be modeled by neural sequence
or latent-variable models. VQ-VAE established vector-quantized latent variables
as a useful discrete interface for images, speech, and generation
\citep{oord2017vqvae}, and later visual tokenizers such as VQGAN made discrete
image latents practical for high-resolution synthesis \citep{esser2021taming}.

\paragraph{Visual tokenizers.}
Image and video tokenizers usually compress visual data into continuous latents
or discrete code grids for downstream generative models. Joint image-video
systems such as OmniTokenizer demonstrate that related visual modalities can
share a tokenizer while retaining useful reconstruction quality
\citep{wang2024omnitokenizer}. Video tokenizers such as MAGVIT and MAGVIT-v2
show that spatio-temporal tokenizers and common image-video vocabularies can be
effective interfaces for video generation and visual language modeling
\citep{yu2022magvit,yu2024magvitv2}. Wavelet-based image tokenizers further
show that multi-scale transforms can be competitive with purely patch-based
visual representations \citep{zhu2024wavelet}. These systems motivate shared
visual token spaces, but they generally keep the token grid dense and remain
within the image-video domain.

\paragraph{Wavelet-space visual tokenization.}
Cosmos Tokenizer is an especially relevant point of comparison: it combines
image/video tokenization, Haar wavelet transforms, autoencoding, and discrete
quantization for efficient visual representations
\citep{nvidia2025cosmos}. This validates the practical value of wavelet-space
visual tokenizers. The present work takes a different target. Rather than
building another image/video wavelet autoencoder, the proposed schema treats
wavelet coefficients as structured tokens whose scale, location, and subband are
part of the token interface. This framing makes it possible to ask whether the
same token schema can cover 1D audio, 2D images, and 3D spacetime video.

\paragraph{Wavelets and multi-scale signal structure.}
Wavelet transforms provide localized multi-scale analysis
\citep{mallat1989theory}. They decompose a signal into low-frequency structure
and higher-frequency detail subbands, making them attractive for natural signals
whose information density is uneven across space and time. Classical image
compression standards such as JPEG 2000 exploit this structure through wavelet
transforms, bit-plane coding, and progressive rate allocation
\citep{taubman2002jpeg2000}. In WAT, these properties are used not only for
compression, but also to define the structure of the token stream.

\paragraph{Audio codec tokenizers.}
Neural audio codec tokenizers compress high-rate waveforms into short discrete
sequences while retaining perceptual quality. SoundStream and EnCodec combine
convolutional encoder-decoder models with residual vector quantization for
high-fidelity neural audio compression \citep{zeghidour2021soundstream,
defossez2022encodec}. Recent systems such as WavTokenizer show that discrete
acoustic tokens can support audio language modeling and generation
\citep{ji2024wavtokenizer}. These methods are typically specialized for audio
and operate on waveform or time-frequency representations. WAT instead uses
audio as the 1D case of a broader signal-field formulation, so that audio
coefficients can share a schema with image and video coefficients.

\paragraph{Quantization and discrete vocabularies.}
The present paper deliberately keeps tokens continuous, but discrete token ids
are a central goal for future work. Vector quantization and residual vector
quantization are standard choices in visual and audio tokenizers, while finite
scalar quantization offers a simpler fixed-code alternative that can reduce
codebook collapse and training complexity \citep{mentzer2024fsq}. These methods
will be needed before WAT can report code utilization, entropy, and true
reconstruction-per-bit.

\paragraph{Adaptive token allocation.}
Dense token grids spend the same number of tokens on smooth regions and complex
regions. Classical compression methods avoid this by allocating more bits to
informative coefficients and fewer bits to predictable parts of the signal. A
wavelet representation makes this allocation explicit: coarse low-frequency
tokens can cover broad smooth structure, while fine-scale tokens can be reserved
for edges, transients, textures, and motion. This motivates comparing fixed-rate
tokenizers against adaptive sparse token selection under matched reconstruction
or bitrate budgets.

\paragraph{Evaluation metrics.}
PSNR and MSE remain useful for controlled rate-distortion measurements, but
learned visual tokenizers are usually evaluated with perceptual and distributional
metrics as well. LPIPS measures perceptual similarity using deep feature
distances \citep{zhang2018lpips}, FID evaluates image distribution quality
\citep{heusel2017fid}, and FVD extends distributional evaluation to generated
video \citep{unterthiner2018fvd}. These metrics motivate the future benchmark
plan in Section~\ref{sec:limitations} and the conclusion.

\paragraph{3D fields and future extension.}
3D Gaussian Splatting and neural fields represent scenes as continuous or
renderable fields rather than dense voxel grids \citep{kerbl20233dgs}.
Compression methods for these representations use pruning, quantization, and
residual vector quantization to reduce storage while preserving rendering
quality \citep{lee2024compact3dgs}. Although the first WAT experiments focus on
audio, images, and video, the field view $x: \Omega \rightarrow \mathbb{R}^C$
is intended to extend naturally to 3D scene attributes, Gaussian parameters,
neural fields, tri-planes, or sparse octree-like parameterizations.

\section{Comparisons}
\label{sec:comparisons}

The closest existing systems overlap with WAT along different axes, but none of
them has the same target: a wavelet-structured token schema that treats audio,
images, and spacetime video as different ranks of a common signal field. This
section is a conceptual comparison rather than a claim of reconstruction
superiority, since the experiments in this paper use a small-scale continuous
autoencoding protocol rather than a production-scale bitrate benchmark.

Classical transform codecs provide the oldest point of contact. JPEG 2000
\citep{taubman2002jpeg2000}, for example, already uses wavelet subbands as an
efficient representation for images. Its goal, however, is hand-engineered
image coding through bit-plane and entropy-coding machinery. WAT instead treats
the wavelet coefficient tuple as a neural token interface whose same fields can
describe one-dimensional audio, two-dimensional images, and three-dimensional
video tensors.

Learned visual tokenizers such as VQ-VAE and VQGAN
\citep{oord2017vqvae,esser2021taming} established the practical importance of
discrete latent grids for generative modeling. Their codes are learned from
data, but the code index usually does not expose an explicit signal-processing
grammar such as scale, subband, or normalized position. WAT makes that grammar
part of the token itself before adding stronger quantization or generative
modeling layers.

Modern audio codecs, including SoundStream, EnCodec, and WavTokenizer
\citep{zeghidour2021soundstream,defossez2022encodec,ji2024wavtokenizer}, are
more mature systems for waveform reconstruction and acoustic tokenization.
They typically use audio-specific convolutional encoders and residual vector
quantization. In contrast, WAT treats audio as the rank-one instance of the same
schema used for images and videos. This is a weaker engineering choice for
domain-specialized compression, but it places cross-modal token structure at
the center of the analysis.

Image-video tokenizers, including OmniTokenizer and the MAGVIT family
\citep{wang2024omnitokenizer,yu2022magvit,yu2024magvitv2}, are closer in their
ambition to share tokenization across related modalities. They generally remain
inside the visual domain and represent data as learned spatial or spacetime
latent grids. WAT differs by starting from an invertible wavelet transform and
asking whether images and videos are merely the rank-two and rank-three cases of
a broader signal-token grammar that also includes audio.

The closest visual reference is Cosmos Tokenizer \citep{nvidia2025cosmos},
which also uses a Haar wavelet front-end for image and video tokenization. The
main distinction is the research question: Cosmos is a production-oriented
visual tokenizer, whereas WAT studies whether wavelet coefficient metadata can
be exposed as a shared schema across signal ranks. In this sense, audio is not
an auxiliary downstream modality but a first-class test of the schema.

Finally, compression methods for Gaussian splatting and radiance-field
representations \citep{kerbl20233dgs,lee2024compact3dgs} address a different
but relevant frontier: compact 3D scene storage. They operate on scene
parameters through pruning, quantization, and residual coding rather than on a
unified audio/image/video token language. They are nevertheless important future
comparisons if WAT is extended from video tensors to continuous 3D fields.

This comparison suggests two useful boundaries. First, WAT should not be read as
a replacement for specialized codecs at this stage: audio codecs and visual
tokenizers remain more mature production systems within their domains.
Second, WAT is not merely another wavelet visual autoencoder. The research
question is whether the coefficient tuple
\[
(\text{value}, \text{rank}, \text{scale}, \text{subband}, \text{position})
\]
can become a common token interface for natural signals, with modality adapters,
sparse rate allocation, and eventual discrete coding layered on top.

\section{Method}

Each modality is formulated as a sampled field
$x: \Omega \rightarrow \mathbb{R}^C$, where $\Omega$ is time for audio, a
two-dimensional spatial grid for images, and a spacetime grid for video. The
model keeps the transform deliberately simple: each input
is mapped through a one-level separable Haar DWT, converted to a shared token
sequence, encoded by a shared continuous token model, scattered back to wavelet
coefficient tensors, and reconstructed with the corresponding Haar IDWT.

\subsection{Wavelet frontend}

For a modality with rank $d \in \{1,2,3\}$ and channel count $C$, a one-level
Haar transform produces $2^d$ subbands. The coefficient tensor has shape
\[
  B \times (C\,2^d) \times g_1 \times \cdots \times g_d ,
\]
where $B$ is the batch size and $g_j$ are the downsampled grid dimensions. Audio
therefore has two subbands, images have four subbands, and video has eight
subbands. The subband order is the low/high channel order produced by the
separable Haar implementation and is preserved exactly by the token flattening
and unflattening path.

\subsection{Shared token schema}

The DWT coefficients are flattened into a common dense schema. For each subband
and grid location, the token value stores the channel coefficients:
\[
  V \in \mathbb{R}^{B \times N \times C}, \qquad
  N = 2^d \prod_{j=1}^{d} g_j .
\]
The same schema is used for all modalities. Each token also carries metadata
\[
  (\text{modality}, \text{rank}, \text{scale}, \text{subband},
  \text{position}) ,
\]
where modality ids are audio, image, and video; rank is 1, 2, or 3; scale is
fixed to zero in the one-level transform; subband ids run from $0$ to
$2^d-1$; and position is a normalized coordinate in $[0,1]^3$. The coordinate
axes are always ordered as $(t,y,x)$; unused axes are filled with zero, so audio
uses $(t,0,0)$ and images use $(0,y,x)$.

This explicit metadata is the main distinction between the proposed schema and
a plain dense latent grid. The model receives not only coefficient values, but
also the grammar that states what each coefficient means.

\subsection{Shared continuous token model}

The autoencoder has small modality-specific value adapters and a shared token
trunk. For modality $m$, the input coefficient value $v_i$ is optionally scaled
by a positive scalar $s_m$ and projected to a common token width:
\[
  e_i = A^{\mathrm{in}}_m(s_m v_i)
      + E_{\mathrm{mod}}(m_i)
      + E_{\mathrm{rank}}(r_i)
      + E_{\mathrm{scale}}(a_i)
      + E_{\mathrm{subband}}(b_i)
      + P(p_i).
\]
Here $A^{\mathrm{in}}_m$ is a linear adapter from the modality channel count to
the shared token width, $E$ terms are learned embeddings for discrete metadata,
and $P$ is a linear position projection from three normalized coordinates.

The shared trunk is token-wise:
\[
  z_i = F_{\mathrm{enc}}(e_i), \qquad
  h_i = F_{\mathrm{dec}}(z_i), \qquad
  \hat{v}_i = A^{\mathrm{out}}_m(h_i) / s_m .
\]
In the present model, $F_{\mathrm{enc}}$ is a LayerNorm-MLP that maps
the token width to a continuous latent dimension, and $F_{\mathrm{dec}}$ maps the
latent token back to the shared token width. No attention, token mixer, or
FSQ/RVQ is used. This isolates the question of whether the shared schema itself
is trainable.

The training loss is signal-space mean squared error,
\[
  \mathcal{L} = \|x - \hat{x}\|_2^2,
\]
while coefficient MSE is logged as a diagnostic. Since the Haar transform is
perfectly invertible in the no-model path, schema round-trip tests verify that
tokenization and scattering introduce no numerical error beyond floating-point
precision.

\subsection{Audio value scaling}

Initial shared-model experiments showed substantially worse audio reconstruction
than image or video reconstruction under the same schema. The failure was traced
to coefficient scale imbalance rather than to an inherent modality-specific
architectural limitation. The model therefore includes a simple value-scaling
parameter before the audio value adapter:
\[
  \tilde{v}_i = s_{\mathrm{audio}} v_i .
\]
The best-performing shared run in the reported grid uses
$s_{\mathrm{audio}}=4$ and keeps image and video scales at one. Per-sample RMS
normalization was also evaluated as an ablation, but the fixed audio scale was
more stable in these experiments.

\subsection{Fixed-rate token selection}

To test whether wavelet coefficients provide a useful allocation signal, the
method includes a non-parametric fixed-rate selection path. For token value $v_i \in
\mathbb{R}^C$, token energy is
\[
  q_i = \frac{1}{C}\sum_{c=1}^{C} v_{i,c}^2 .
\]
Given a keep ratio $\rho$, a selector constructs a boolean mask $M \in
\{0,1\}^{B \times N}$ and dropped tokens are set to zero:
\[
  v_i^{\mathrm{masked}} = M_i v_i .
\]
The masked coefficient values are scattered back to the DWT tensor and decoded
with the same Haar IDWT. The comparison includes global energy top-$k$,
per-subband energy top-$k$, uniform stride, random, and a lowpass-first
baseline. This path has no
learned parameters and directly tests whether coefficient energy is a useful
cross-modal token importance signal.

\subsection{Masked shared training}

The masked autoencoder uses the same dense token layout, but the model receives
masked token values as input and is trained to reconstruct the full signal. This
keeps the schema unchanged while measuring whether the shared trunk can infer
missing coefficients from sparse wavelet observations. Results are reported for
\texttt{metadata=none} and \texttt{metadata=full} because earlier matched-rate
experiments showed that additive metadata is not uniformly beneficial.

\section{Experiments}

The experiments evaluate whether the shared schema can reconstruct real audio,
image, and video data with one shared continuous token trunk.

\subsection{Data and experimental configuration}

The evaluation uses three public datasets: Speech Commands for audio
\citep{warden2018speechcommands}, EuroSAT RGB for images
\citep{helber2019eurosat}, and DAVIS 2017 for video
\citep{ponttuset2017davis}. Audio clips are sampled as mono waveforms with
16,384 samples. Images are resized to $64 \times 64$. Video samples use 8 RGB
frames at $64 \times 64$.

All reported results are measurements from a compact shared model
trained with round-robin modality updates. The separate baseline uses one small
wavelet autoencoder per modality. PSNR is computed from MSE with data range
1.0; image and video tensors are in $[0,1]$, while audio waveforms are int16
samples normalized by 32768.

\subsection{Rate proxies and statistical scope}

Because the present model uses continuous latents rather than discrete codes,
this paper reports rate proxies rather than true bitrate. The dense token
counts per sample are 16,384 for audio, 4,096 for images, and 32,768 for video.
For dense autoencoders, continuous latent scalar budgets are compared, defined
as token count times latent dimension. For fixed-rate sparse experiments, the
reported quantities are kept-token ratio and kept-token count. These quantities
measure representation size under the proposed schema, but they do not include
codebook entropy, index coding overhead, entropy modeling, or
bits-per-pixel/sample.

The learned autoencoder results are single-run measurements and should
be read as trend evidence rather than final statistical estimates. The
non-parametric fixed-rate selection sweep is repeated over selector seeds for
stochastic baselines, but the trained dense and masked models do not yet report
mean and standard deviation across training seeds. Larger-scale follow-up
experiments are needed to replace these proxies with explicit compression ratios
and multi-seed rate-distortion curves.

\subsection{Main results}

Table~\ref{tab:main-results} compares a separate modality baseline, a shared
schema model with unit value scales, and a shared schema model with
$s_{\mathrm{audio}}=4$. The unit-scale shared model already improves image and
video over the separate baseline, but audio degrades relative to the audio-only
model. Scaling audio coefficients before the shared value adapter recovers audio
quality while preserving the visual gains.

\begin{table}[ht]
\centering
\caption{Preliminary validation reconstruction quality. Higher PSNR is better
and lower MSE is better.}
\label{tab:main-results}
\small
\begin{tabular}{lrrrrrr}
\toprule
Model & A-MSE & A-PSNR & I-MSE & I-PSNR & V-MSE & V-PSNR \\
\midrule
Separate baseline
  & $8.34{\times}10^{-5}$ & 40.79
  & $5.11{\times}10^{-3}$ & 22.91
  & $8.11{\times}10^{-3}$ & 20.91 \\
Shared schema, unit scale
  & $2.35{\times}10^{-3}$ & 26.30
  & $9.65{\times}10^{-4}$ & 30.15
  & $5.58{\times}10^{-3}$ & 22.53 \\
Shared schema, audio scale 4
  & $1.02{\times}10^{-4}$ & 39.92
  & $1.16{\times}10^{-3}$ & 29.37
  & $4.04{\times}10^{-3}$ & 23.93 \\
\bottomrule
\end{tabular}
\end{table}

The best-performing shared result in Table~\ref{tab:main-results} is obtained
with $s_{\mathrm{audio}}=4$. Compared
with the separate baseline, this model is within 0.87 dB on audio, improves
image PSNR by 6.46 dB, and improves video PSNR by 3.03 dB. These numbers should
not be interpreted as final compression quality, because the model remains dense
and continuous. They nevertheless indicate that a single shared wavelet token
grammar can represent all three modalities in this autoencoding setting.

\subsection{Matched-rate dense sweep}

The primary confound is capacity: the shared dense schema has one latent
vector per wavelet coefficient token, while the separate baseline uses latent
channels on a coefficient grid. The evaluation therefore compares against a
matched separate baseline whose latent channels are set to $2^d$ times the
shared latent dimension. Table~\ref{tab:matched-rate} shows the result at shared
latent dimension 16.

\begin{table}[ht]
\centering
\caption{Matched continuous latent scalar budget at shared latent dimension 16.
The metadata-free shared model often outperforms additive metadata on visual
modalities.}
\label{tab:matched-rate}
\begin{tabular}{lrrr}
\toprule
Model & Audio PSNR & Image PSNR & Video PSNR \\
\midrule
Separate matched & 43.93 & 22.47 & 21.76 \\
Shared full metadata & 39.92 & 29.37 & 23.93 \\
Shared no metadata & 41.29 & 30.41 & 29.42 \\
\bottomrule
\end{tabular}
\end{table}

\begin{figure}[t]
\centering
\includegraphics[width=\linewidth]{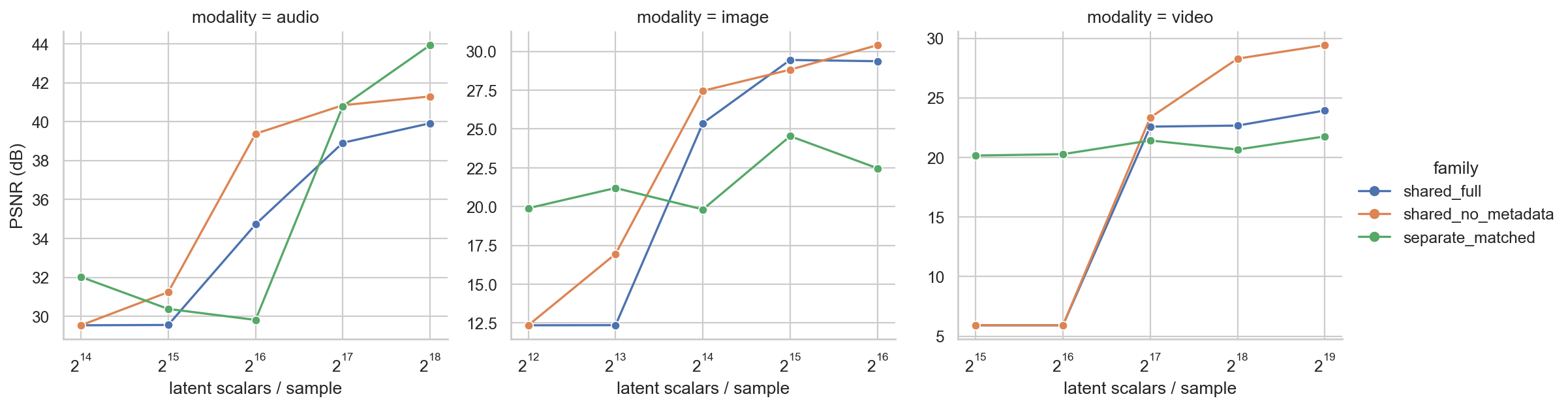}
\caption{Rate-distortion sweep. The shared schema remains favorable on image and
video under matched continuous latent scalar budgets, while audio
continues to expose a modality-specific normalization and modeling gap.}
\label{fig:rate-distortion}
\end{figure}

This sweep changes the interpretation of the shared result. The image and video
advantage is not simply a larger continuous bottleneck, because the separate
baseline is matched by scalar latent count. At the same time, explicit additive
metadata is not the source of the gain: removing metadata improves image and
video in this dense setting. Figure~\ref{fig:rate-distortion} shows the same
trend across the reported rate sweep.

\subsection{Audio scaling ablation}

Table~\ref{tab:audio-scaling} reports the 150-step ablation that motivated
the audio scale. Audio improves sharply when moving from unit scale to scale 2
or 4, while image and video remain stable. Increasing the scale to 8 starts to
degrade audio, suggesting that the relevant factor is not merely coefficient
magnitude but a moderate alignment of coefficient ranges before the shared
trunk. Per-sample RMS normalization helps audio relative to unit scale but
reduces image quality in this setting.

\begin{table}[ht]
\centering
\caption{Validation PSNR for audio value-scaling ablations.}
\label{tab:audio-scaling}
\begin{tabular}{lrrr}
\toprule
Setting & Audio PSNR & Image PSNR & Video PSNR \\
\midrule
Unit scale & 27.56 & 23.62 & 21.53 \\
Audio scale 2 & 30.36 & 24.96 & 21.62 \\
Audio scale 4 & 30.81 & 24.98 & 21.62 \\
Audio scale 8 & 28.95 & 24.57 & 21.61 \\
Sample RMS normalization & 29.15 & 22.87 & 21.71 \\
\bottomrule
\end{tabular}
\end{table}

\subsection{Fixed-rate energy selection}

The next experiment evaluates sparse token allocation without learning. For each
batch, DWT token values are masked according to a fixed keep ratio and decoded
directly through the Haar IDWT. Table~\ref{tab:selection-deltas} reports average
PSNR gains over uniform and random selection, excluding the trivial 100\% keep
ratio.

\begin{table}[ht]
\centering
\caption{Average PSNR gain of \texttt{energy\_global} selection over fixed-rate
baselines across keep ratios 50\%, 25\%, 10\%, 5\%, and 1\%.}
\label{tab:selection-deltas}
\begin{tabular}{lrrr}
\toprule
Modality & vs. uniform & vs. random & Minimum gain vs. uniform \\
\midrule
Audio & +16.73 & +16.73 & +2.17 \\
Image & +16.90 & +16.92 & +0.15 \\
Video & +15.86 & +15.85 & +0.85 \\
\bottomrule
\end{tabular}
\end{table}

\begin{figure}[t]
\centering
\includegraphics[width=\linewidth]{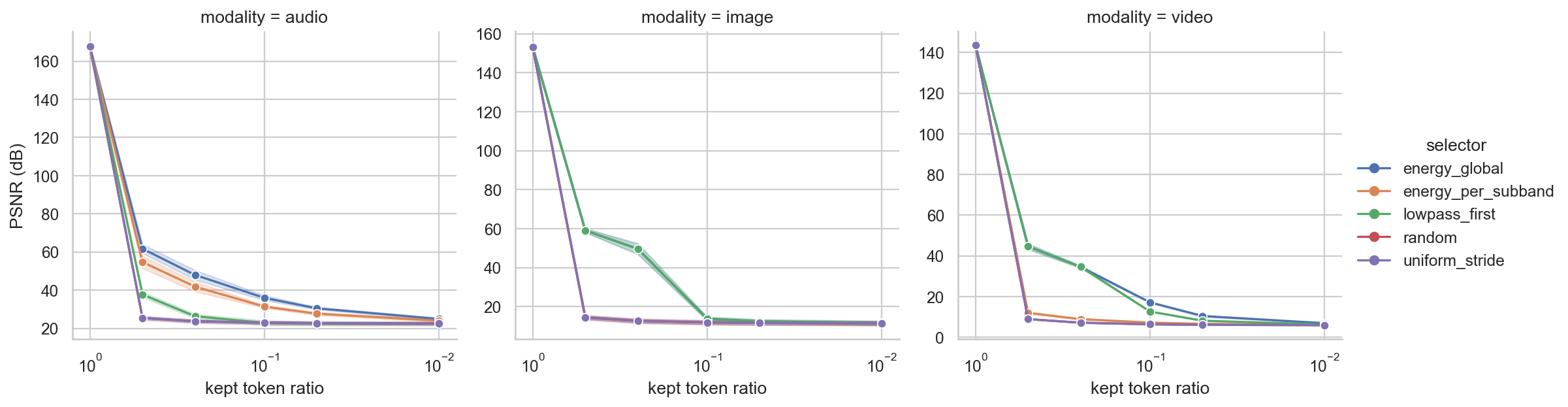}
\vspace{0.4em}
\includegraphics[width=\linewidth]{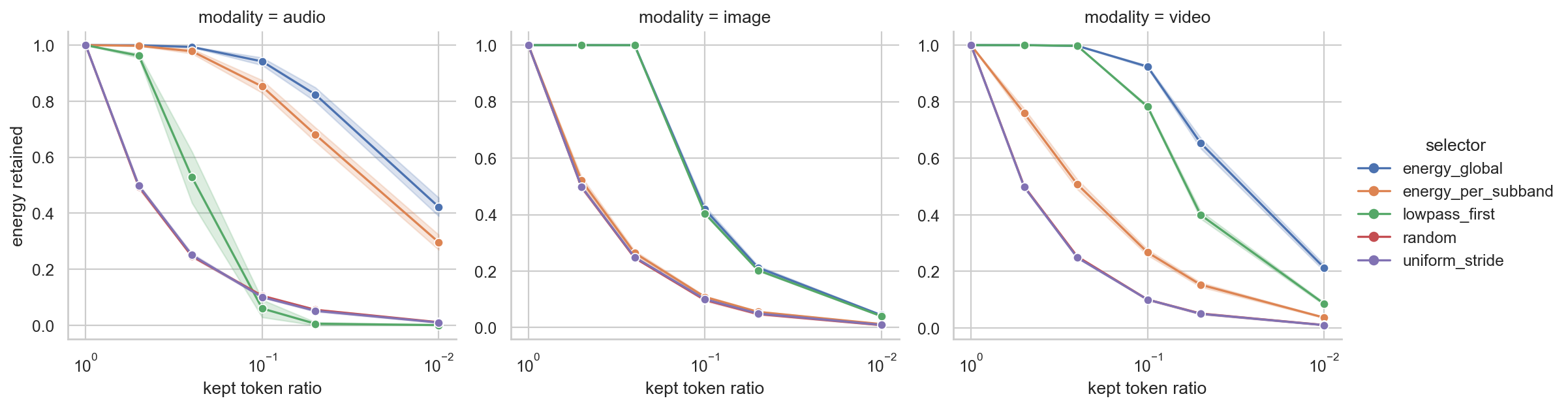}
\caption{Non-parametric fixed-rate token selection. Top: reconstruction PSNR
across keep ratios and selectors. Bottom: retained wavelet energy. Energy-based
selection consistently preserves more reconstruction quality than uniform or
random fixed-rate token layouts.}
\label{fig:fixed-rate-selection}
\end{figure}

Global energy selection outperforms the baselines on all three modalities and
every compressed keep ratio. Per-subband energy selection is helpful for audio
but much weaker for image and video. A lowpass-first baseline nearly matches global energy for
images, is close but weaker for video, and is substantially worse for audio.
This suggests that visual reconstruction is dominated by low-frequency structure, while audio
requires cross-subband energy allocation. Figure~\ref{fig:fixed-rate-selection}
visualizes both reconstruction quality and retained coefficient energy.

\subsection{Masked sparse training}

Given the strength of non-parametric energy selection, the shared autoencoder is
also trained with masked token values as input and full reconstruction as the
target. Table~\ref{tab:masked-training} lists the best masked
results from the evaluated grid over metadata mode, latent dimension, and keep
ratio.

\begin{table}[ht]
\centering
\caption{Best masked sparse shared results. The video result uses only 50\% of
dense tokens and substantially exceeds the dense shared video baseline.}
\label{tab:masked-training}
\begin{tabular}{llrrr}
\toprule
Modality & Best setting & Keep ratio & PSNR & MSE \\
\midrule
Audio & none, latent 8 & 0.50 & 32.60 & $5.50{\times}10^{-4}$ \\
Image & full, latent 8 & 0.50 & 29.98 & $1.01{\times}10^{-3}$ \\
Video & none, latent 16 & 0.50 & 34.45 & $3.59{\times}10^{-4}$ \\
\bottomrule
\end{tabular}
\end{table}

\begin{figure}[t]
\centering
\includegraphics[width=\linewidth]{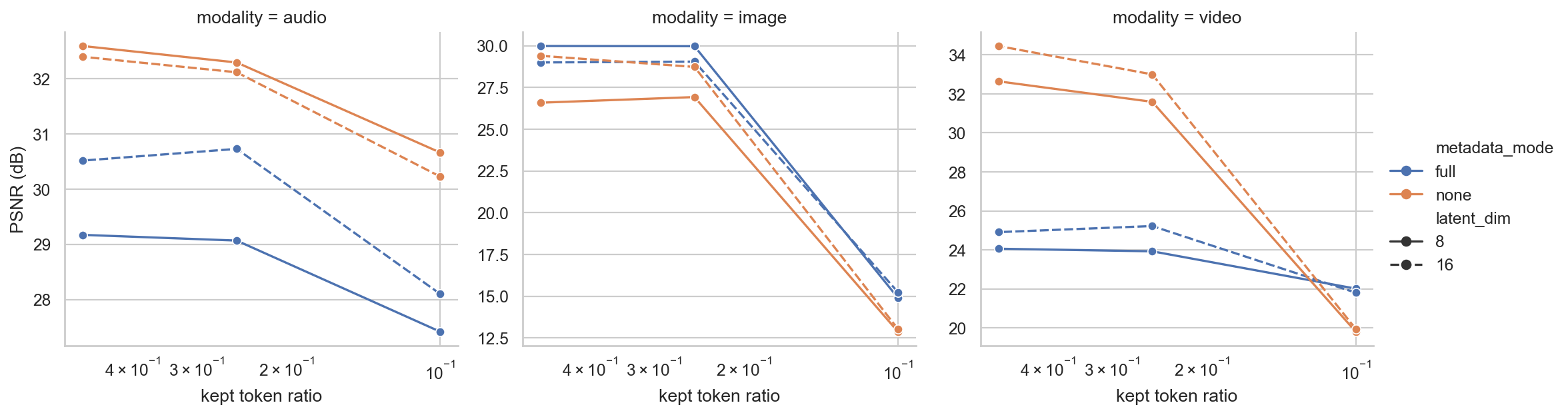}
\caption{Masked sparse shared training. The curves compare metadata modes,
latent dimensions, modalities, and fixed keep ratios under the energy-based
masking configuration. The best-performing sparse result appears for video at a 50\%
keep ratio without additive metadata.}
\label{fig:masked-training}
\end{figure}

The metadata effect is modality- and rate-dependent. Audio is better with no
metadata in every masked setting. Video is much better without metadata at 25\%
and 50\% keep ratios, but full metadata helps at 10\%. Images often benefit from
full metadata, especially at latent dimension 8. These results indicate that
metadata should remain an ablation rather than the default explanation for the
value of unified tokenization. Figure~\ref{fig:masked-training} summarizes the
sparse training curves.

\FloatBarrier

\subsection{Interpretation}

These results support four conclusions. First, the shared token schema
is numerically sound: DWT coefficients can be flattened into a single token
grammar and scattered back exactly in unit tests. Second, a small shared
token-wise model can learn real audio, image, and video reconstruction without
attention or modality-specific trunks. Third, coefficient normalization matters:
audio is an equal participant in the dense schema only after its value scale is
calibrated. Fourth, wavelet energy is an effective cross-modal sparse allocation
signal in the non-parametric selection experiments, and sparse masked training
is promising for video in this small-scale setting.

These claims remain bounded by the scale and statistical scope of the experiments.
Section~\ref{sec:limitations} separates these limitations from the positive
evidence.

\section{Limitations}
\label{sec:limitations}

This paper is intentionally framed as an early-stage empirical study. The
reported experiments support the feasibility of a shared wavelet token schema,
but they do not establish a production-quality tokenizer.

\paragraph{Prototype scale.}
The experiments use low-resolution images, short video clips, a compact
token-wise MLP trunk, and reconstruction losses centered on MSE. This is useful
for isolating the schema question, but it is not comparable to production-scale
visual tokenizers trained with larger datasets, higher resolutions, stronger
spatio-temporal architectures, perceptual losses, and multi-stage optimization.
Absolute PSNR values should therefore be interpreted as small-scale
autoencoding measurements, not state-of-the-art reconstruction results.

\paragraph{Rate proxy rather than bitrate.}
All reported rates are proxies: dense token count, continuous latent scalar
count, or fixed keep ratio. Since the model has no FSQ, VQ, RVQ, entropy model,
or learned rate controller, it does not report true bitrate, bits per pixel,
bits per audio sample, or reconstruction-per-bit. The matched-rate comparisons
are still informative for capacity control, but they do not replace a full
rate-distortion benchmark.

\paragraph{Statistical strength.}
The learned autoencoder tables and curves are single-run measurements. They
reveal consistent qualitative trends, such as the importance
of audio coefficient scaling and the strength of energy-based sparse selection,
but they do not provide confidence intervals or training-seed variance. The
non-parametric selection results are less sensitive because they do not train a
model, but future learned results should report mean and variance across seeds.

\paragraph{Continuous tokens only.}
The present study uses continuous latent tokens. It has no discrete vocabulary,
codebook utilization measurement, entropy estimate, dead-code analysis, or
downstream generative model consuming the tokens. The evidence therefore
supports a unified schema and sparse token interface, not a universal discrete
token vocabulary.

\paragraph{Metadata is unresolved.}
Explicit metadata is part of the proposed token grammar, but the additive
embedding implementation evaluated here is not uniformly beneficial. It helps
some image settings while degrading audio and several video settings. This
suggests that metadata should be treated as a conditioning design problem, possibly via
gating, FiLM-style modulation, routing, or selector-only use, rather than as an
already validated source of improvement.

\section{Future Work}
\label{sec:future-work}

\paragraph{Directions for scaling.}
Several extensions are needed before the proposed schema can be evaluated as a
practical tokenizer. First, evaluation should move to comparable rate-distortion
protocols: DAVIS and TokenBench for video, MS-COCO or ImageNet for images,
explicit spatial-temporal compression ratios such as $4{\times}8{\times}8$ and
$8{\times}16{\times}16$, and metrics including PSNR, SSIM, LPIPS, rFID, and
rFVD. Second, the one-level Haar front-end should be extended to a multi-level
wavelet pyramid where scale is an active token field, not a fixed placeholder.
Third, the token-wise MLP should be replaced or augmented with local
convolutional decoders, causal and non-causal spatio-temporal token mixers, and
attention or state-space modules that can model context across neighboring
coefficients. Fourth, sparse token allocation should move from fixed-rate energy
selection to learned rate control, entropy-aware token budgets, and matched
sparse separate baselines, so that gains can be measured at a real bitrate
rather than only by kept-token count.

Fifth, the continuous latent should be discretized only after the sparse setting
is better understood. FSQ is a useful first candidate because it tends to train robustly;
RVQ or VQ variants should then be compared with code utilization, entropy,
dead-code rate, and reconstruction-per-bit as primary diagnostics. Sixth,
training should include stronger reconstruction objectives: L1 and perceptual
losses for image detail, temporal or optical-flow losses for video consistency,
audio-aware spectral or psychoacoustic losses for waveform quality, and
adversarial fine-tuning only after the rate-distortion behavior is understood.
Finally, the unified-token claim should remain broader than visual
reconstruction alone: audio should remain in the benchmark, and the same
scale-subband-position grammar should be tested on continuous 3D fields such as
Gaussian-splat attributes, neural fields, tri-planes, or sparse scene grids.

Together, these directions would test a broader hypothesis: natural signals may
share a common multi-scale token grammar, while modality-specific adapters,
learned rate allocation, and discrete coding specialize that grammar for audio,
image, video, and 3D tokenization tasks.

\section{Conclusion}

This paper investigates wavelet-inspired tokenization as a bridge between
classical signal compression and modern neural tokenizers. It presents a shared
schema for audio, images, and video: all three modalities are converted to the
same coefficient-value plus metadata grammar, processed by one shared continuous
token trunk, and reconstructed through modality-specific Haar IDWT paths.

The experiments support the feasibility of a unified wavelet token schema and
sparse token interface in a controlled small-scale autoencoding setting. With a
simple audio coefficient scale, the dense shared model is close to the small
audio-only baseline and improves over the small visual baselines used here.
Matched-rate sweeps suggest that the visual gains are not explained only by
bottleneck size. Fixed-rate energy selection is an effective cross-modal
allocation signal, and masked sparse training gives encouraging video results.

The evidence also narrows the claim. Additive metadata embeddings are
not a universal source of improvement: they help some image settings but often
degrade audio and video. The results therefore support a unified schema and sparse
token interface, not yet a universal discrete vocabulary. Within the limitations
described above, the evidence motivates further evaluation of the schema, sparse
selection mechanism, and tokenizer architecture.

\section*{Statements}

The author used AI-assisted writing tools for language editing, organization,
and drafting support during preparation of this manuscript. The author reviewed
and edited the resulting text and takes full responsibility for the content,
claims, experiments, and conclusions.

\clearpage
\bibliographystyle{unsrtnat}
\bibliography{references}

@article{zhu2024wavelet,
  title = {Wavelet-Based Image Tokenizer for Vision Transformers},
  author = {Zhu, Zhenhai and Soricut, Radu},
  journal = {arXiv preprint arXiv:2405.18616},
  year = {2024}
}

@article{wang2024omnitokenizer,
  title = {OmniTokenizer: A Joint Image-Video Tokenizer for Visual Generation},
  author = {Wang, Junke and Jiang, Yi and Yuan, Zehuan and Peng, Binyue and Wu, Zuxuan and Jiang, Yu-Gang},
  journal = {arXiv preprint arXiv:2406.09399},
  year = {2024}
}

@article{nvidia2025cosmos,
  title = {Cosmos World Foundation Model Platform for Physical AI},
  author = {{NVIDIA}},
  journal = {arXiv preprint arXiv:2501.03575},
  year = {2025}
}

@article{ji2024wavtokenizer,
  title = {WavTokenizer: An Efficient Acoustic Discrete Codec Tokenizer for Audio Language Modeling},
  author = {Ji, Shengpeng and Jiang, Ziyue and Wang, Wen and Chen, Yifu and Fang, Minghui and Zuo, Jialong and Yang, Qian and Cheng, Xize and Wang, Zehan and Li, Ruiqi and others},
  journal = {arXiv preprint arXiv:2408.16532},
  year = {2024}
}

@article{lee2024compact3dgs,
  title = {Compact 3D Gaussian Splatting for Static and Dynamic Radiance Fields},
  author = {Lee, Joo Chan and Rho, Daniel and Sun, Xiangyu and Ko, Jong Hwan and Park, Eunbyung},
  journal = {arXiv preprint arXiv:2408.03822},
  year = {2024}
}

@article{warden2018speechcommands,
  title = {Speech Commands: A Dataset for Limited-Vocabulary Speech Recognition},
  author = {Warden, Pete},
  journal = {arXiv preprint arXiv:1804.03209},
  year = {2018}
}

@article{helber2019eurosat,
  title = {EuroSAT: A Novel Dataset and Deep Learning Benchmark for Land Use and Land Cover Classification},
  author = {Helber, Patrick and Bischke, Benjamin and Dengel, Andreas and Borth, Damian},
  journal = {IEEE Journal of Selected Topics in Applied Earth Observations and Remote Sensing},
  volume = {12},
  number = {7},
  pages = {2217--2226},
  year = {2019},
  publisher = {IEEE}
}

@article{ponttuset2017davis,
  title = {The 2017 DAVIS Challenge on Video Object Segmentation},
  author = {Pont-Tuset, Jordi and Perazzi, Federico and Caelles, Sergi and Arbel{\'a}ez, Pablo and Sorkine-Hornung, Alexander and Van Gool, Luc},
  journal = {arXiv preprint arXiv:1704.00675},
  year = {2017}
}

@inproceedings{oord2017vqvae,
  title = {Neural Discrete Representation Learning},
  author = {van den Oord, Aaron and Vinyals, Oriol and Kavukcuoglu, Koray},
  booktitle = {Advances in Neural Information Processing Systems},
  volume = {30},
  year = {2017}
}

@inproceedings{esser2021taming,
  title = {Taming Transformers for High-Resolution Image Synthesis},
  author = {Esser, Patrick and Rombach, Robin and Ommer, Bj{\"o}rn},
  booktitle = {Proceedings of the IEEE/CVF Conference on Computer Vision and Pattern Recognition},
  pages = {12873--12883},
  year = {2021}
}

@article{yu2022magvit,
  title = {MAGVIT: Masked Generative Video Transformer},
  author = {Yu, Lijun and Cheng, Yong and Sohn, Kihyuk and Lezama, Jos{\'e} and Zhang, Han and Chang, Huiwen and Hauptmann, Alexander G. and Yang, Ming-Hsuan and Hao, Yuan and Essa, Irfan and Jiang, Lu},
  journal = {arXiv preprint arXiv:2212.05199},
  year = {2022}
}

@inproceedings{yu2024magvitv2,
  title = {Language Model Beats Diffusion---Tokenizer is Key to Visual Generation},
  author = {Yu, Lijun and Lezama, Jos{\'e} and Gundavarapu, Nitesh B. and Versari, Luca and Sohn, Kihyuk and Minnen, David and Cheng, Yong and Birodkar, Vighnesh and Gupta, Agrim and Gu, Xiuye and Hauptmann, Alexander G. and Gong, Boqing and Yang, Ming-Hsuan and Essa, Irfan and Ross, David A. and Jiang, Lu},
  booktitle = {International Conference on Learning Representations},
  year = {2024}
}

@inproceedings{mentzer2024fsq,
  title = {Finite Scalar Quantization: VQ-VAE Made Simple},
  author = {Mentzer, Fabian and Minnen, David and Agustsson, Eirikur and Tschannen, Michael},
  booktitle = {International Conference on Learning Representations},
  year = {2024}
}

@article{mallat1989theory,
  title = {A Theory for Multiresolution Signal Decomposition: The Wavelet Representation},
  author = {Mallat, Stephane G.},
  journal = {IEEE Transactions on Pattern Analysis and Machine Intelligence},
  volume = {11},
  number = {7},
  pages = {674--693},
  year = {1989},
  publisher = {IEEE}
}

@article{taubman2002jpeg2000,
  title = {{JPEG2000}: Standard for Interactive Imaging},
  author = {Taubman, David S. and Marcellin, Michael W.},
  journal = {Proceedings of the IEEE},
  volume = {90},
  number = {8},
  pages = {1336--1357},
  year = {2002},
  publisher = {IEEE}
}

@article{zeghidour2021soundstream,
  title = {SoundStream: An End-to-End Neural Audio Codec},
  author = {Zeghidour, Neil and Luebs, Alejandro and Omran, Ahmed and Skoglund, Jan and Tagliasacchi, Marco},
  journal = {arXiv preprint arXiv:2107.03312},
  year = {2021}
}

@article{defossez2022encodec,
  title = {High Fidelity Neural Audio Compression},
  author = {D{\'e}fossez, Alexandre and Copet, Jade and Synnaeve, Gabriel and Adi, Yossi},
  journal = {arXiv preprint arXiv:2210.13438},
  year = {2022}
}

@article{kerbl20233dgs,
  title = {3D Gaussian Splatting for Real-Time Radiance Field Rendering},
  author = {Kerbl, Bernhard and Kopanas, Georgios and Leimk{\"u}hler, Thomas and Drettakis, George},
  journal = {ACM Transactions on Graphics},
  volume = {42},
  number = {4},
  pages = {139:1--139:14},
  year = {2023},
  publisher = {ACM}
}

@inproceedings{zhang2018lpips,
  title = {The Unreasonable Effectiveness of Deep Features as a Perceptual Metric},
  author = {Zhang, Richard and Isola, Phillip and Efros, Alexei A. and Shechtman, Eli and Wang, Oliver},
  booktitle = {Proceedings of the IEEE Conference on Computer Vision and Pattern Recognition},
  pages = {586--595},
  year = {2018}
}

@inproceedings{heusel2017fid,
  title = {GANs Trained by a Two Time-Scale Update Rule Converge to a Local Nash Equilibrium},
  author = {Heusel, Martin and Ramsauer, Hubert and Unterthiner, Thomas and Nessler, Bernhard and Hochreiter, Sepp},
  booktitle = {Advances in Neural Information Processing Systems},
  volume = {30},
  year = {2017}
}

@article{unterthiner2018fvd,
  title = {Towards Accurate Generative Models of Video: A New Metric and Challenges},
  author = {Unterthiner, Thomas and van Steenkiste, Sjoerd and Kurach, Karol and Marinier, Raphael and Michalski, Marcin and Gelly, Sylvain},
  journal = {arXiv preprint arXiv:1812.01717},
  year = {2018}
}

\clearpage
\appendix
\section{Experimental Details}
\label{app:experimental-details}

\subsection{Data Preprocessing}
The experiments use Speech Commands v0.02, EuroSAT RGB, and DAVIS 2017
train/validation data. Audio examples are converted to mono waveforms of 16,384
samples and normalized by 32768. Images are resized to $64 \times 64$ RGB
tensors in $[0,1]$. Video examples use 8 RGB frames resized to $64 \times 64$.
Training batches are sampled independently for each modality, and validation
batches are sampled from the corresponding validation split.

\subsection{Shared Autoencoder Configuration}
Unless otherwise stated, the shared model uses token width 32, latent dimension
16, hidden dimension 64, AdamW optimization with learning rate $10^{-3}$, batch
size 2, and 300 training steps. The three modalities are trained with
round-robin updates. Validation is run every 50 steps, and reported validation
values are averaged over two validation batches per modality. The default value
scales are $s_{\mathrm{audio}}=4$, $s_{\mathrm{image}}=1$, and
$s_{\mathrm{video}}=1$, with no per-sample RMS normalization unless explicitly
noted. The default metadata mode for the metadata-free shared model is
\texttt{metadata=none}; the full-metadata ablation uses additive modality, rank,
scale, subband, and position embeddings.

\subsection{Separate and Matched-Rate Baselines}
The separate baseline trains one wavelet autoencoder per modality. For
matched-rate comparisons, the separate model's latent channel count is set to
$2^d$ times the shared latent dimension for a modality of rank $d$, matching the
number of continuous latent scalars per sample. The rate-distortion sweep uses
shared latent dimensions $1,2,4,8,16$ and seed 0. Reported rates are continuous
latent scalar counts rather than entropy-coded bitrates.

\subsection{Fixed-Rate Token Selection}
The non-parametric selection experiment evaluates six keep ratios: full density,
50\%, 25\%, 10\%, 5\%, and 1\%. The compared selectors are
\texttt{energy\_global}, \texttt{energy\_per\_subband},
\texttt{uniform\_stride}, \texttt{random}, and \texttt{lowpass\_first}. The
selection sweep uses seeds 0, 1, and 2 for stochastic selectors and four
validation batches per modality. Token energy is the mean squared coefficient
value across channels. Dropped tokens are set to zero before inverse wavelet
reconstruction.

\subsection{Masked Sparse Training}
Masked sparse training uses \texttt{energy\_global} selection, input-zero
masking, keep ratios $0.5,0.25,0.1$, latent dimensions 8 and 16, and metadata
modes \texttt{none} and \texttt{full}. Each run uses the same optimizer,
training length, value scales, and validation protocol as the dense shared
autoencoder unless otherwise stated.

\subsection{Metrics and Scope}
PSNR is computed from signal-space MSE with data range 1.0. Coefficient MSE is
logged as a diagnostic but is not used as the training objective. The reported
learned-model results are single-seed measurements; they are intended to
characterize trends in the proposed schema rather than final compression
performance.

\end{document}